\documentclass[conference]{IEEEtran}

\usepackage{algorithmicx}
\usepackage{algorithm}
\usepackage[noend]{algpseudocode}
\usepackage{cite}
\usepackage{epsfig}
\usepackage{color}
\usepackage{graphics}
\usepackage{amssymb,eepic}
\usepackage[cmex10]{amsmath}
\usepackage{mathtools}
\usepackage{proof}
\usepackage{tikz}
\usepackage{url}

\usetikzlibrary{arrows}
\usetikzlibrary{positioning}
\usetikzlibrary{automata}

\ifCLASSINFOpdf
\else
\fi

\newtheorem{definition}{Definition}
\newtheorem{theorem}{Theorem}
\newtheorem{corollary}{Corollary}
\newtheorem{lemma}{Lemma}

\usepackage[font=footnotesize,caption=false]{subfig}

\usepackage{hyperref}

\newcommand{\ioco}{\textbf{ioco}}
\newcommand{\uioco}{\textbf{uioco}}
\newcommand{\tioco}{\textbf{tioco}}
\newcommand{\cspio}{\textbf{cspio}}
\newcommand{\sut}{\textsc{SUT}}
\newcommand{\iolts}{\textsc{IOLts}}
\newcommand{\riolts}{\textsc{R-IOLts}}
\newcommand{\qinit}{\hat{q}}
\newcommand{\traces}[1]{\textbf{Traces}(#1)}
\newcommand{\straces}[1]{\textbf{Straces}(#1)}
\newcommand{\after}{\textbf{after}}

\newcommand{\Labels}{L}
\newcommand{\InputLabels}{L^{I}}
\newcommand{\OutputLabels}{L^{O}}
\newcommand{\out}[2]{\textbf{out}_{#2}(#1)}  

\newcommand{\intcomp}[2]{{#1} \otimes {#2}}	

\newcommand{\projection}[2]{#1_{\downarrow #2}}

\newcommand{\envdet}[1]{E_{\textrm{max}}(#1)} 

\newcommand{\hide}[2]{h_{#2}(#1)}	        
\newcommand{\fhide}[2]{\hat{h}_{#2}(#1)}
\newcommand{\money}{\emph{coin?}}
\newcommand{\utee}{\emph{utee?}}

\newcommand{\ucoffee}{\emph{ucoffee?}}
\newcommand{\umilk}{\emph{umilk?}}
\newcommand{\outmtee}{\emph{mtee!}}

\newcommand{\outmcoffee}{\emph{mcoffee!}}
\newcommand{\outmcoffeemilk}{\emph{mcoffeemilk!}}
\newcommand{\indone}{\emph{done?}}
\newcommand{\msg}{\emph{msg!}}

\newcommand{\inmtee}{\emph{mtee?}}
\newcommand{\inmcoffee}{\emph{mcoffee?}}
\newcommand{\inmcoffeemilk}{\emph{mcoffeemilk?}}
\newcommand{\outdone}{\emph{done!}}

\newcommand{\coffee}{\emph{coffee!}}
\newcommand{\coffeemilk}{\emph{coffeemilk!}}

\newcommand{\gmoney}{\emph{coin}}
\newcommand{\mtee}{\emph{mtee}}
\newcommand{\mcoffee}{\emph{mcoffee}}
\newcommand{\mcoffeemilk}{\emph{mcoffeemilk}}
\newcommand{\done}{\emph{done}}
\newcommand{\gucoffee}{\emph{ucoffee}}
\newcommand{\gutee}{\emph{utee}}
\newcommand{\gcoffee}{\emph{coffee}}
\newcommand{\gcoffeemilk}{\emph{coffeemilk}}
\newcommand{\gumilk}{\emph{umilk}}

\begin{document}
\title{Compositional Specifications for $\ioco$ Testing}

\author{\IEEEauthorblockN{Przemys\l aw Daca \\ and Thomas A. Henzinger}
\IEEEauthorblockA{IST Austria\\
Klosterneuburg, Austria \\ 
\{przemek, tah\}@ist.ac.at}
\and
\IEEEauthorblockN{Willibald Krenn \\ and Dejan Ni\v{c}kovi\'{c}}
\IEEEauthorblockA{AIT Austrian Institute of Technology GmbH.\\
Vienna, Austria \\
\{willibald.krenn, dejan.nickovic\}@ait.ac.at}}


%


\maketitle

%
\IEEEpeerreviewmaketitle

\begin{abstract}
Model-based testing is a promising technology for black-box software and hardware testing, in which test cases are generated automatically from high-level specifications. Nowadays, systems typically consist of multiple interacting components and, due to their complexity, testing presents a considerable portion of the effort and cost in the design process. Exploiting the compositional structure of system specifications can considerably reduce the effort in model-based testing. Moreover, inferring properties about the system from testing its individual components allows the designer to reduce the amount of integration testing.

In this paper, we study compositional properties of the $\ioco$-testing theory. We propose a new approach to composition and hiding operations, inspired by contract-based design and interface theories. These operations preserve behaviors that are compatible under composition and hiding, and prune away incompatible ones. The resulting specification characterizes the input sequences for which the unit testing of components is sufficient to infer the correctness of component integration without the need for further tests. We provide a methodology that uses these results to minimize integration testing effort, but also to detect potential weaknesses in specifications. While we focus on asynchronous models and the $\ioco$ conformance relation, the resulting methodology can be applied to a broader class of systems.
\end{abstract}
\section{Introduction}
\label{sec:intro}

\begin{figure}[htb]
\begin{center}
\scalebox{0.65}{ 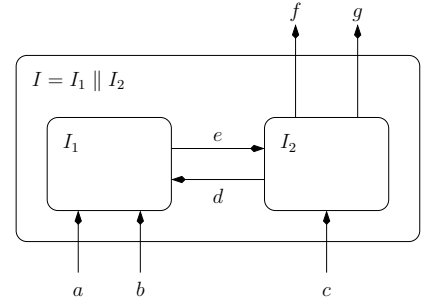 }
\end{center}
\caption{A system consisting of two interacting components.}
\label{fig:embedded}
\end{figure}

Modern software and hardware system design usually involves the integration of interacting components that work together 
in order to realize some requested behavior.  Fig.~\ref{fig:embedded} illustrates two components $I_{1}$ and 
$I_{2}$ that are composed together to form the system $I$. The complexity of the individual components together with their elaborate 
cooperation protocols often results in behavioral faults. Therefore, verification and validation methods are applied 
to ensure that the embedded system satisfies its specification. This is in particular true for safety-critical 
designs, for which correctness evidence is imposed by the regulation bodies (see for example the automotive 
standard ISO 26262~\cite{iso}). 

Up to date, design simulation combined with testing remains the preferred technique 
in industry to demonstrate the correctness of software and hardware systems. Typically, verification engineers need to 
first test individual system components ({\em unit testing} of $I_{1}$ and $I_{2}$), 
and then test the complete system ({\em integration testing} of $I$). 
This process relies on 
verification engineers manually generating test vectors from specifications given as informal (natural language) 
requirements. This process is inherently time consuming, ad-hoc and prone to human errors. As a result, testing 
represents the main bottleneck in the design of complex systems today.

Model-based testing is a technology that provides formalization and  automation 
to the test case generation and execution process, thus reducing time and cost of systems design. 
In model-based 
testing a {\em system-under-test} ($\sut$), denoted by $I$, 
is tested for conformance to its formal specification model $S$, derived from informal requirements. 
While $S$ is a formal object, $I$ is a ``black-box'', a physical implementation with 
unknown internal structure. The $\sut$ $I$ can be accessed by the tester only through its external interface. 
In order to reason about the compliance of $I$ to $S$, one needs to 
use the {\em testing assumption} (see~\cite{tretmans-mbt}), stating that $I$ can be modeled in the same formalism as $S$ and that 
$I$ is {\em receptive} ({\em input-enabled}), i.e. it accepts all inputs at any point in time. In contrast to $I$, $S$ does not need to 
be receptive. Lack of input (\emph{under-specification}) in a given state of the specification  models the assumption that 
the external environment for $I$ does not provide that input. If the environment nevertheless emits this input action, 
the specification allows $I$ to freely choose its response.

In this paper, we focus on the $\ioco$-testing 
theory~\cite{tretmans-mbt}, a model-based testing framework 
for input/output labeled transition systems ($\iolts$). 
The $\ioco$-testing theory is centered around the {\em input/output conformance relation} $\ioco$. 
Informally, we say that an implementation $I$ $\ioco$-conforms to its specification $S$ if any 
experiment derived from $S$ and executed on $I$ leads to an output in $I$ that is foreseen by $S$. In the $\ioco$-testing theory, the 
lack of outputs or internal transitions is observable via a \emph{quiescence} action.

In its original form, $\ioco$-testing does not take the compositional aspects of systems  into account. For example, in
 Fig.~\ref{fig:embedded}, $I$ is the result of composing the components $I_{1}$ and $I_{2}$. Typically, 
the actions over which the two components synchronize ($e$ and $d$ in the example) are 
{\em hidden} and become unobservable to the tester after integration. In order to cope with costly testing of large  
systems, results of unit testing individual components need to be used to infer properties about the composed system, to avoid 
or at least minimize expensive integration tests. The compositional $\ioco$-testing  
problem can be formulated as follows: if $I_{1}$ $\ioco$-conforms to its specification model $S_{1}$ and $I_{2}$ $\ioco$-conforms to its 
specification model $S_{2}$, can we infer that $I$ also $\ioco$-conforms to $S$, where $I$ and $S$ denote the composed implementation and 
specification after hiding synchronization actions?

This question was first addressed in~\cite{tretmans-comp}. The authors show that in the general case 
neither parallel composition of specifications nor hiding of 
actions are compositional in the $\ioco$-testing theory. This result is not surprising. Parallel composition 
is an operation tailored to receptive models. We argue that it is not an appropriate operator for composing under-specified 
models where one component can generate an output which is not expected as an input by the other.  In addition, the hiding operation introduces 
partial observation over the actual state of the $\sut$ and can result in confusion regarding the under-specified parts of the 
system. The authors of~\cite{tretmans-comp} propose two alternative 
restrictions to the specification models in order to preserve compositionality of $\ioco$. The 
first option is to disallow under-specification of inputs. This is a very strong requirement in practice, since components 
are usually designed to operate in constrained environments. The second option allows starting with under-specification, 
but requires {\em demonic completion} 
of specification models --- an operation which makes the assumptions about the component's environment explicit and thus 
makes the specification model input-enabled.
We claim that demonic completion can hide important information from the tester about the poor quality of a specification 
and thus obscure its original intent.

We propose a different approach to compositional $\ioco$-testing which, in contrast to~\cite{tretmans-comp},  
does not restrict the specification models. We define two operations --- {\em friendly composition} and 
{\em hiding} --- that are tailored 
to the integration of non-receptive specifications. They are based on a game-theoretic {\em optimistic} approach, inspired 
by interface theories~\cite{intf1,intf2}. The result of the friendly 
composition is the overall specification that integrates the component 
specifications while pruning away any inputs that lead to incompatible interactions between the components. The friendly hiding operation 
prunes away inputs that lead to states which are ambiguous with respect to under-specification after hiding. 
After composing the component specifications followed by hiding of synchronizing actions, 
the resulting specification defines all input sequences for which no integration testing is 
needed --- the correct integration follows from the conformance of the individual components to their specification. 
In addition to these technical results, this paper provides guidelines to identify specifications that are poorly modeled 
for compositional testing. 
We argue that pruned input sequences often indicate weaknesses in the specification 
and can be addressed by: (1) strengthening the specifications; 
(2) making more outputs observable; and (3) integration testing. Indeed, the proper formalization of requirements 
resulting in high-quality component specifications is crucial for exploiting the compositional nature of systems in 
testing. Investing efforts in improving models can considerably minimize expensive integration tests. 
We discuss methodological aspects of using our technical results to improve component models and to tailor them to 
compositional testing. 

In Section~\ref{sec:motivating}, we further motivate the problem of compositional testing with $\ioco$ and provide an informal 
overview of our approach, illustrating it with a vending machine example. We also identify modeling issues and discuss problems related to 
compositional $\ioco$-testing and sketch possible solutions. Section~\ref{sec:prelim} recalls the basics of the $\ioco$-testing theory including  
the known results about compositional $\ioco$-testing. We provide the formal presentation of our approach in Section~\ref{sec:interface} and 
evaluate it in Section~\ref{sec:eval}. We present related work in more detail in Section~\ref{sec:related}, and finally conclude the 
paper in Section~\ref{sec:conclusion}, giving future perspectives for our work. The proofs are presented in a separate 
report~\cite{tr}.

\section{Overview and Motivating Example}
\label{sec:motivating}

In this section, we develop the example of a {\em drink vending machine}, which we use to 
(1) further motivate the problem; (2) highlight the difficulties of compositional testing within the $\ioco$-theory; and 
(3) provide an informal overview of our proposed approach to tackle the problem.

\begin{figure}[htb]
\centering
\scalebox{0.5}{ 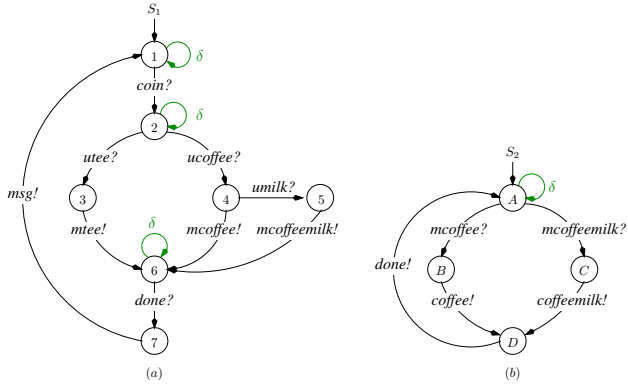 }
\caption{Vending machine --- specification of components: (a) user interface $S_{1}$; and (b) drink maker $S_{2}$.}
\label{fig:vending-spec}
\end{figure}

The drink vending machine consists of the {\em user interface} and the {\em drink maker} components. 
The user interface specification $S_{1}$ is shown in Fig.~\ref{fig:vending-spec}~(a). $S_{1}$ 
requires that the user first inserts a coin ($\money$)\footnote{We consistently use the symbol $\emph{?}$ to 
denote an input, and the symbol $\emph{!}$ to denote an output action.}, and then selects either a tee ($\utee$) or 
a coffee ($\ucoffee$). After choosing the coffee, the user can also request milk ($\umilk$) for the coffee. The 
drink request ($\outmtee$, $\outmcoffee$ or $\outmcoffeemilk$) 
is forwarded to the drink maker, and the user interface waits for an acknowledgment ($\indone$) that the 
drink was delivered to the user. When the drink is ready, the user interface emits a message ($\msg$) to the user and returns to its 
initial state. 

The drink maker specification $S_2$,  
depicted in Fig.~\ref{fig:vending-spec}~(b), waits for a drink request 
($\inmcoffee$ or $\inmcoffeemilk$) from the user interface. 
Upon receiving the drink request, $S_{2}$ signals the delivery of the drink to the user (actions 
$\coffee$ and $\coffeemilk$) and finally an acknowledgment ($\outdone$) is sent to the user interface. 
Note that $S_2$ is under-specified in its initial state $A$ --- it omits 
the action $\inmtee$ and thus makes an assumption that the user interface never requests a tee.

We note that in 
states $1$ and $2$ of $S_{1}$, and state $A$ of $S_{2}$ only inputs are allowed. In the $\ioco$-testing theory, 
such states are called {\em quiescent}, where the quiescence denotes the absence of observable outputs and internal actions.
The absence of outputs is considered to be observable, and is marked as a special $\delta$ action (green self-loops in 
Fig.~\ref{fig:vending-spec}). Since quiescence is usually not explicitly modeled by the designer,  
we omit marking quiescent transitions in the rest of the section.

\begin{figure}[htb]
\centering
\scalebox{0.5}{ 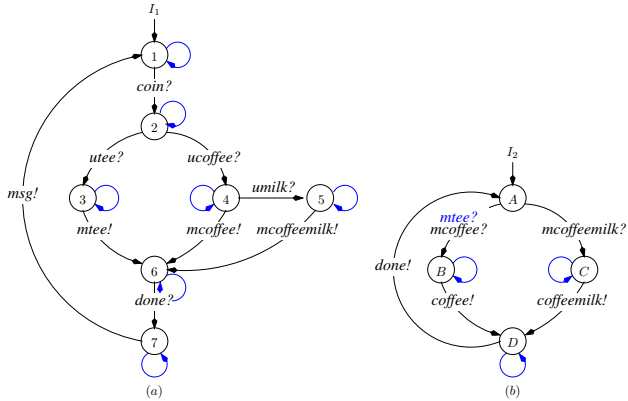 }
\caption{Vending machine --- implementation of components: (a) user interface $S_{1}$; and (b) drink maker $S_{2}$. To improve readability 
of the figure, we 
omit the missing input labels on blue self-loops. For instance, state $1$ of $S_{1}$ is labeled with input actions $\utee$, $\ucoffee$, $\umilk$ and 
$\indone$, while states $B$, $C$ and $D$ of $S_{2}$ are labeled with $\inmtee$, $\inmcoffee$ and $\inmcoffeemilk$.}
\label{fig:vending-impl}
\end{figure}

Specifications $S_1$ and $S_2$ give a certain freedom in implementing user interface and drink machine components. In particular, 
implementations can choose how to treat under-specified (unexpected) inputs. For instance, $S_{1}$ assumes that the user 
inserts a coin before choosing the drink. If the user swaps the order of actions, and first orders a drink, $S_1$ allows the implementation 
to react to this input in an arbitrary way. Fig.~\ref{fig:vending-impl} depicts possible implementations $I_1$ and $I_2$ of 
their respective specifications $S_1$ and $S_2$. 
The user interface implementation $I_1$ 
closely follows its specification and silently ignores all unexpected inputs (marked in blue). For example, 
if the user requests a coffee before inserting a 
coin (in state $1$), this request is silently consumed by $I_{1}$. This implementation is 
$\ioco$-conformant to its specification because it never generates an output which is not foreseen by $S_1$. Similarly to $I_1$, the 
drink maker implementation $I_2$ also silently consumes unexpected inputs in all states, except in the state $A$. In the initial state, 
$I_{2}$ reacts to a tee request ($\inmtee$) by moving to the state $B$, 
from which a coffee ($\coffee$) is delivered to the user. Although preparing a 
coffee upon a tee request may not be a logical behavior, it is
$\ioco$-conformant to $S_2$ --- the specification does not impose any particular reaction to the tee request in its initial state, giving to 
correct implementations complete freedom of handling such input. 

\begin{figure}[htb]
\centering
\scalebox{0.5}{ 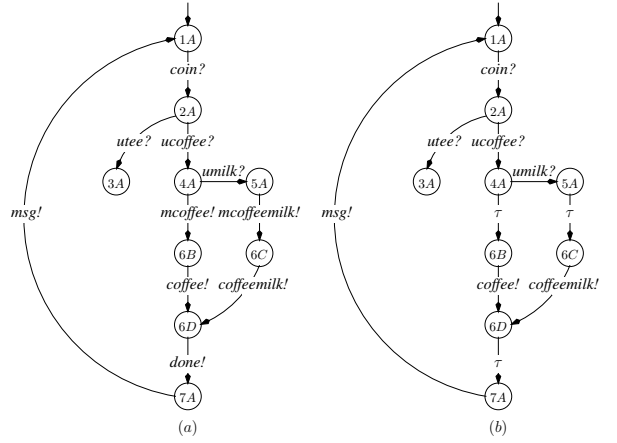 }
\caption{Vending machine --- specification: (a) parallel composition of $S_{1}$ and $S_{2}$; (b) with shared actions 
$\mtee$, $\mcoffee$, $\mcoffeemilk$ and $\done$ hidden.}
\label{fig:vending-spec-comp}
\end{figure}

\begin{figure}[htb]
\centering
\scalebox{0.5}{ 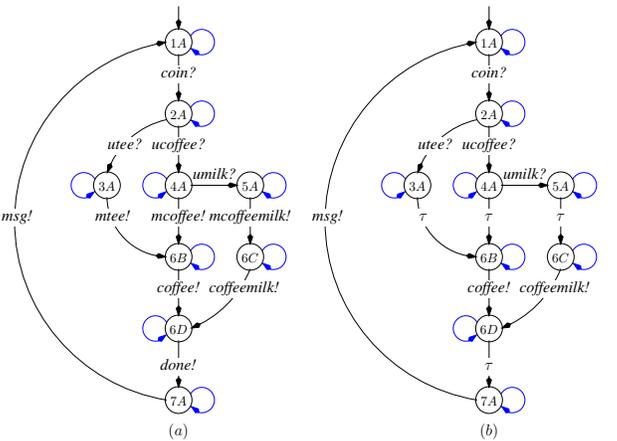 }
\caption{Vending machine --- implementation: (a) parallel composition of $I_{1}$ and $I_{2}$; (b) with shared actions 
$\mtee$, $\mcoffee$, $\mcoffeemilk$ and $\done$ hidden.}
\label{fig:vending-impl-comp}
\end{figure}

In the classic $\ioco$-theory~\cite{tretmans-comp}, specifications are combined with the {\em parallel composition} 
operation for input/output transitions systems~\cite{lynch}. Informally, parallel composition of two specifications is their Cartesian product, where each 
specification is allowed to take local actions independently, but the two specifications must synchronize on shared actions. 
Fig.~\ref{fig:vending-spec-comp}~(a) depicts the parallel composition of $S_{1}$ and $S_{2}$, where 
$\mtee$, $\mcoffee$, $\mcoffeemilk$ and $\done$ are the shared actions. 
In addition to parallel composition, we may wish to {\em hide} shared actions, which are often only used to synchronize components, but 
are not observable by the external user.
In the vending machine example, the user can observe inputs to the vending machine (inserting a coin and choosing the drink) and 
the outputs from the machine (the actual drink and the acknowledgment message). 
The actions $\mtee$, $\mcoffee$, $\mcoffeemilk$ and $\done$ are 
used for proper synchronization between the user interface and the drink maker components and are not visible to the user. 
Fig.~\ref{fig:vending-spec-comp}~(b) depicts the parallel composition of specifications $S_{1}$ and $S_{2}$ in which the shared actions 
are hidden (denoted by the special action $\tau$). Fig.~\ref{fig:vending-impl-comp}~(a) and (b) show the parallel composition of the 
component implementations without and with hiding of the synchronization actions, respectively.

While  $I_{1}$ is $\ioco$-conformant to $S_{1}$ and $I_{2}$ is 
$\ioco$-conformant to $S_{2}$, the composition of $I_{1}$ and $I_{2}$ is not 
$\ioco$-conformant to the composition of $S_{1}$ and $S_{2}$ (with or without hiding of shared actions), as shown in~\cite{tretmans-comp}. 
We now explain the reasons for non-compositionality of $\ioco$ in the vending machine example.

Assuming that shared actions are not hidden, consider 
a test case starting with the input sequence $\money \cdot \utee$. According to the specification (Fig.~\ref{fig:vending-spec-comp}~(a)), 
the only allowed observation after reading this sequence is quiescence, since state $3A$ does not have any outgoing transitions labeled by an 
output action. However, the composed implementation  (Fig.~\ref{fig:vending-impl-comp}~(a)) emits $\outmtee$, an output action 
not allowed by the specification after reading the same sequence. It follows that $\ioco$-conformance is not preserved by 
parallel composition. 

Parallel composition is an operation tailored to combining receptive components  --- whenever a 
component outputs a shared action the other component is by definition able to consume it from any of its local states. This is not the 
case for non-receptive models. Indeed, $S_{1}$ emits the shared action 
$\outmtee$ in its local state $3$, while the drink maker specification $S_{2}$ is not ready to consume it in its local state $A$, i.e. 
the assumption of $S_2$ is not fulfilled by $S_1$. This 
results in a ``deadlock'' state $3A$ in the parallel composition of $S_1$ and $S_2$. However, the intended meaning of under-specifying 
the action $\inmtee$ in the 
state $A$ of $S_{2}$ is that $S_2$ is free to choose any reaction to this unexpected action.  This is in contrast to 
what happens in state $3A$ of the composed specification.

We now consider the case when the shared actions are hidden. After reading the input sequence 
$\money \cdot \ucoffee$, the state of the composed system after hiding is not uniquely defined --- 
it can be either $4A$ or $6B$  (Fig.~\ref{fig:vending-impl-comp}~(b)) since the user cannot observe whether 
the hidden action $\outmcoffee$ has taken place. Note that the specification (Fig.~\ref{fig:vending-spec-comp}~(b)) 
leaves the input $\umilk$ unspecified in $6B$, but not in $4A$, thus resulting in an ambiguity on what an external observer 
can expect as the reaction to this input. 
In fact, according to the $\ioco$-theory, the only allowed observable output after executing the sequence $\money \cdot \ucoffee \cdot \umilk$ 
is $\coffeemilk$, while the composed implementation can output both $\coffeemilk$ (if in state $6C$) or $\coffee$ (if in state $6B$).

\begin{figure}[htb]
\centering
\scalebox{0.5}{ 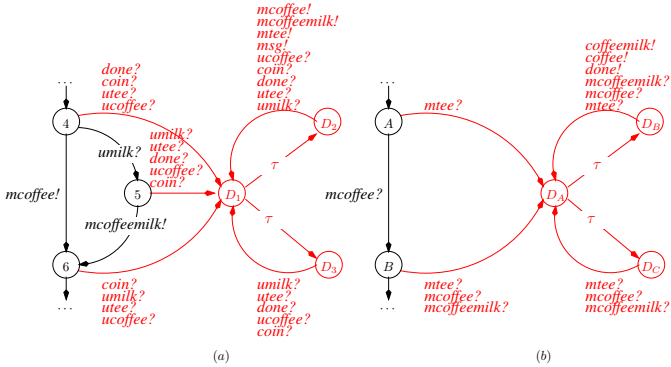 }
\caption{Parts of the demonically completed specifications: (a) $S_{1}$; and (b) $S_{2}$. States and transitions resulting from the completion are 
marked in red.}
\label{fig:dc}
\end{figure}

In~\cite{tretmans-comp}, the authors propose two solutions to the above anomalies. Both solutions guarantee that 
$\ioco$-conformance is preserved by parallel composition and hiding. The first solution requires specification models to be receptive. 
We claim that this restriction is too strong --- under-specification of inputs is one of 
the most powerful modeling tools for specifying open systems. A component is almost always expected to work correctly only in constrained 
contexts, and under-specification of inputs allows to exactly define a valid operating environment. The second solution allows 
non-receptive models, but requires their {\em demonic completion} --- an operation that makes the model effectively input-enabled. 
Demonic completion results in adding transitions labeled with the under-specified inputs from every state to a newly inserted ``sink'' portion of the graph.
The sink portion essentially self-loops with all inputs and outputs. Demonic completion makes  
the intended meaning of input under-specification explicit:  
when a system receives an unexpected input, it has the full freedom to choose its reaction to this input. Fig.~\ref{fig:dc} depicts parts 
of the demonically completed specifications $S_{1}$ and $S_{2}$. 

While demonic completion preserves the 
intended meaning of specifications, we argue that it does not provide a fully satisfactory solution to compositional $\ioco$-testing. 
First, the resulting specification after demonic completion  
increases in size. Although linear, this increase is still important for extensively under-specified models. 
For instance, $S_{1}$ has $7$ states and $9$ transitions, while its size increases 
to $10$ states and $55$ transitions after completion. 
Second, demonic completion obfuscates the distinction between foreseen and unspecified interactions between components. 
In Fig.~\ref{fig:dc}~(a) the information that the input action $\ucoffee$ is not expected in state $6$ is lost. 

This lack of distinction between foreseen and unexpected interactions between components masks the fact that we often deal with component 
specifications of poor quality. This results in a composition which does not faithfully represent the intended behavior of the overall system. 
In particular, composition with hiding of demonically 
completed specifications may result in many vacuous behaviors. We illustrate this problem with the 
example from Fig.~\ref{fig:dc}. After composing demonically completed variants of $S_{1}$ and $S_{2}$, and hiding the synchronization action $\mcoffee$, the external observer cannot distinguish between states $4A$ and $6B$, which in contrast to the original composition 
(see Fig.~\ref{fig:vending-spec-comp}~(b)), now both admit the input action $\umilk$. However, $\umilk$ triggers a transition from $6B$ to the 
state $D_{1}B$, from which the demonically completed specification $S_{1}$ allows all possible behaviors, including the one in which the 
acknowledgment message is sent back to the user without any drink being served.

\begin{figure}[htb]
\centering
\scalebox{0.5}{ 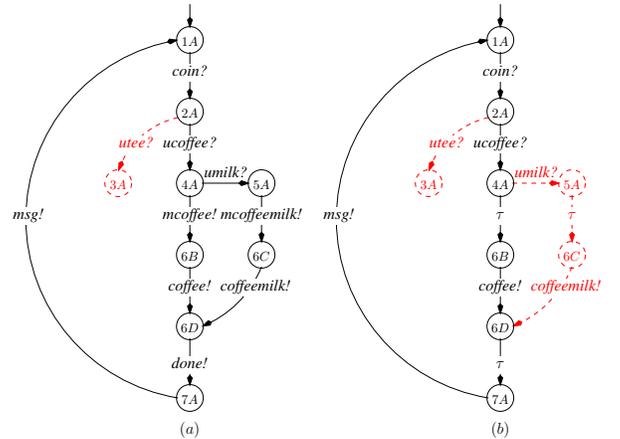 }
\caption{Vending machine --- overall specification: (a) friendly composition of $S_{1}$ and $S_{2}$; (b) with friendly 
hiding of synchronization actions 
$\mtee$, $\mcoffee$, $\mcoffeemilk$ and $\done$.}
\label{fig:vm-spec-intf}
\end{figure}

We propose an alternative approach to composition and hiding that provides additional information to the designer. 
It is inspired by  
interface theories~\cite{intf1, intf2}. We provide a new composition operation, called the {\em friendly composition}, which takes an 
{\em optimistic} approach to combining two specifications. Following the optimistic approach, two specifications are compatible for 
composition, if there exists {\em some} context in which they can interact while both satisfying their guarantees. We have seen in 
Fig.~\ref{fig:vending-spec-comp}~(a) that the interaction of two specifications (user interface and drink maker) 
results in states in which one component is allowed to emit an output which is not expected by the other component 
(action $\mtee!$ in the state $3A$). We declare such states as {\em ambiguous states}, and compute the {\em maximal} compound 
environment which avoids such states.
The algorithm that computes the friendly composition of two specifications 
prunes all states from which the environment cannot prevent the composed system from reaching an ambiguous state. The resulting 
composite specification combines the compatible interactions of two component specifications. 
Fig.~\ref{fig:vm-spec-intf}~(a) depicts the friendly composition of 
$S_{1}$ and $S_{2}$, where red dashed transitions and states 
are the ones pruned away from their parallel composition. 

Similarly to friendly composition, we define the {\em friendly hiding} operation, which prunes away from the specification with 
hidden actions all the states that can become ambiguous when interacting with an external environment. The resulting 
specification is depicted in Fig.~\ref{fig:vm-spec-intf}~(b), where red dashed transition and states are pruned away from the 
composite specification of $S_{1}$ and $S_{2}$ after hiding. 

The main technical contribution of this paper is the definition of the friendly composition and friendly hiding operations, for which we show that 
they preserve $\ioco$-conformance. In contrast to combining demonic completion with parallel composition and hiding, our approach results 
in composite specifications of smaller size. The resulting composite specification defines behaviors for which no integration testing is needed. Apart 
from the technical contribution, we argue that friendly composition and hiding expose weaknesses of 
component specifications to the designer. The pruned behaviors after applying friendly composition and hiding indicate that assumptions made by 
individual component specifications may be too weak and deserve more careful analysis. We claim that the time spent on improving the quality of the component specifications so that they allow compositional testing is rewarded with the avoidance of integration tests.
We distinguish between the following scenarios 
in using our approach to derive better compositional specifications.

\noindent {\em Scenario A:} the designer can guarantee that the composed system will be used in the context defined by the assumptions of the friendly-composed 
specification and that the ambiguous interactions will never take place. In this case, no additional integration testing is needed. In the vending 
machine example, this would mean that the designer has the possibility of disabling the tee and milk request buttons.

\noindent {\em Scenario B:} by hiding  shared actions information about the internal state of the $\sut$ may be lost, resulting in ambiguous states. Better observability can be achieved by keeping some shared actions visible to the external environment.
From the technical point of view, keeping the $\outmcoffee$ action visible in the vending machine specification allows for 
compositional testing. 

\noindent {\em Scenario C:} the designer cannot guarantee that the composed system 
will be used in the context defined by the friendly-composed specification 
assumptions. In this case, the specification is too weak and needs a revision. In our example, 
the race between the input action $\umilk$ and the output action $\outmcoffee$ in $S_1$ indicates a poor specification. 
We propose a different specification, which requires an additional action and in which the user is expected 
to make all requests before the machine is able to process them.

\section{Preliminaries}
\label{sec:prelim}

In this section, we define labeled transition systems, parallel composition and hiding operations, 
input/output conformance relation ($\ioco$), and recall the previous results on compositional properties of 
$\ioco$.

\subsection{Labeled Transition Systems}
\label{sec:lts}
An {\em input/output labeled transition system} ($\iolts$) is a formal model for specifying reactive systems. 
An  $\iolts$ $A$ is a tuple $(Q, \InputLabels, \OutputLabels, T, \qinit)$, where $Q$ is a countable set of \emph{states},
$\InputLabels$ and $\OutputLabels$ are disjoint countable sets of \emph{input} and \emph{output labels}, $\qinit \in Q$ is the \emph{initial state} and $T$ is the \emph{transition relation}. We denote by $\Labels = \InputLabels \cup \OutputLabels$ the set of all labels of an $\iolts$. To avoid ambiguity we may use subscripts, like $Q_A$, to indicate that an element belongs to an $\iolts$ $A$.
We consider 
$\iolts$ with possibly {\em silent} transitions, denoted by $\tau$, hence the transition relation is defined as 
$T \subseteq Q \times (L \cup \{ \tau \}) \times Q$, with $\tau \not \in \Labels$.  An $\iolts$ is 
said to be {\em receptive}, denoted by $\riolts$, if for all $q \in Q$ and for all $a \in \InputLabels$, 
there exists an outgoing transition from $q$ labeled by $a$. For instance, specifications $S_{1}$ and $S_{2}$ in 
Fig.~\ref{fig:vending-spec} are $\iolts$, while implementations $I_{1}$ and $I_{2}$ in Fig.~\ref{fig:vending-impl} are 
$\riolts$. {\em Strongly-convergent} $\iolts$ are transition systems that do not 
have loops consisting of only silent transitions. 
We use the standard abbreviated notation, where $\mu \in \Labels \cup \{\tau\}$ and $a \in \Labels$
$$
\begin{array}{lcl}
q \xrightarrow{\mu} q' & \equiv & (q, \mu, q') \in T \\
q \xrightarrow{\mu_{1} \cdot \ldots \cdot \mu_{n}} q' & \equiv & \exists q_{0}, \ldots, q_{n} \; \textrm{st.} \; 
q = q_{0} \xrightarrow{\mu_{1}} q_{1} \\
&&\xrightarrow{\mu_{2}} \ldots \xrightarrow{\mu_{n}} q_{n} = q' \\
q \xrightarrow{\mu_{1} \cdot \ldots \cdot \mu_{n}} & \equiv & \exists q' \; \textrm{st.} \; q \xrightarrow{\mu_{1} \cdot \ldots \cdot \mu_{n}} q' \\
q \not \xrightarrow{\mu_{1} \cdot \ldots \cdot \mu_{n}} & \equiv & \neg \exists q' \; \textrm{st.} \; q \xrightarrow{\mu_{1} \cdot \ldots \cdot \mu_{n}} q' \\
q \xRightarrow{\epsilon} q' & \equiv & q = q' \; \textrm{or} \; q \xrightarrow{\tau\cdot \ldots \cdot \tau} q' \\

q \xRightarrow{a} q' & \equiv & \exists q_{1},q_{2} \; \textrm{st.} \; 
q \xRightarrow{\epsilon} q_{1} \xrightarrow{a} q_{2} \xRightarrow{\epsilon} q' \\

q \xRightarrow{a_{1} \cdot \ldots \cdot a_{n}} q' & \equiv & \exists q_{0}, \ldots, q_{n} \; \textrm{st.} \; 
q = q_{0} \xRightarrow{a_{1}} q_{1} \\ 
&&\xRightarrow{a_{2}} \ldots \xRightarrow{a_{n}} q_{n} = q' \\

q \xRightarrow{a_{1} \cdot \ldots \cdot a_{n}} & \equiv & \exists q' \; \textrm{st.} \; q \xRightarrow{a_{1} \cdot \ldots \cdot a_{n}} q' \\

q \not \xRightarrow{a_{1} \cdot \ldots \cdot a_{n}} & \equiv & \neg \exists q' \; \textrm{st.} \; q \xRightarrow{a_{1} \cdot \ldots \cdot a_{n}} q' \\
\end{array}
$$

A sequence $\sigma \in (\Labels \cup \{ \tau\})^{+}$ is an {\em execution} of an $\iolts$ $A$ if $\qinit_A \xrightarrow{\sigma}$. A 
sequence $\sigma \in \Labels^{*}$ is a trace of $A$ if $\qinit_A \xRightarrow{\sigma}$. We denote by $\traces{A}$ the set of all 
traces of $A$. The sequence 
$\gmoney \cdot \gucoffee \cdot \tau  \cdot \gcoffee$ is an execution of the specification shown in Fig.~\ref{fig:vending-spec-comp}~(b), 
while $\gmoney \cdot \gucoffee \cdot \gcoffee$ is its trace. Given a subset of labels 
$\Labels' \subseteq \Labels$ and $\sigma$ a sequence over $\Labels$, we denote by $\projection{\sigma}{\Labels'}$ 
the projection $\sigma$ to the set of labels $\Labels'$.

We say that a state $q \in Q$ of $A$ is {\em quiescent}, denoted by $\delta(q)$, if it has no outgoing output or internal actions. Quiescent states 
emit a special \emph{quiescence} action $\delta$, which indicates that $A$ cannot proceed without input from the environment.
The \emph{suspension automata} are $\iolts$, where quiescent actions are made explicit.
Formally, given an $\iolts$ $A = (Q, \InputLabels, \OutputLabels, T, \qinit)$, its suspension automaton $A_{\delta}$ is 
the $\iolts$ $A_{\delta} = (Q, \InputLabels, \OutputLabels \cup \{\delta\}, T \cup T_{\delta}, \qinit)$, where 
$T_{\delta} = \{q \xrightarrow{\delta} q~|~\delta(q)\}$. We denote by $\straces{A}$ the set
$\{\sigma \in (\Labels \cup \{\delta\})^{*}~|~\qinit \xRightarrow{\sigma}_{A_\delta}\}$ of 
traces of $A_\delta$, also called {\em suspension traces}. Specifications $S_{1}$ and 
$S_{2}$ shown in Fig.~\ref{fig:vending-spec} show explicitly quiescence actions, where for example 
$\delta \cdot \gmoney \cdot \delta$ is in $\straces{S_{1}}$. 

\subsection{Parallel Composition}
\label{sec:comp}

Two components can be integrated if their input/output actions do not conflict. In particular, we require that 
the intersection of their input (output) label sets is empty. Formally, we say that two $\iolts$ $A_{1}$ and $A_{2}$ are
{\em composable} if $\InputLabels_{1} \cap \InputLabels_{2} = \OutputLabels_{1} \cap \OutputLabels_{2} = \emptyset$.
When two composable $\iolts$ are composed, they synchronize on shared actions and move independently on other actions. 
Formally, {\em parallel composition} is defined as follows.

\begin{definition}[Parallel composition]
\label{def:p-composition}
Let $A_{1} = (Q_{1}, \InputLabels_{1}, \OutputLabels_{1}, T_{1}, \qinit_{1})$ and 
$A_{2} = (Q_{2}, \InputLabels_{2}, \OutputLabels_{2}, T_{2}, \qinit_{2})$ be two composable $\iolts$. 
Their {\em parallel composition}, denoted by $A_{1} \parallel A_{2}$, is the $\iolts$ 
$(Q_{1 \parallel 2}, \InputLabels_{1 \parallel 2}, \OutputLabels_{1 \parallel 2}, T_{1 \parallel 2}, \qinit_{1 \parallel 2})$, where
$Q_{1 \parallel 2} = Q_1 \times Q_2$, $\InputLabels_{1 \parallel 2} = (\InputLabels_{1} \backslash \OutputLabels_{2}) \cup 
(\InputLabels_{2} \backslash \OutputLabels_{1})$, $\OutputLabels_{1 \parallel 2} = \OutputLabels_{1} \cup \OutputLabels_{2}$, $\qinit_{1 \parallel 2} = (\qinit_{1}, \qinit_{2})$, 
 and 
$T_{1 \parallel 2}$ is defined by the rules:
\[
  \infer{(q_1, q_2) \xrightarrow{\mu}_{A_{1 \parallel 2}} (q'_1, q_2)}{q_1 \xrightarrow{\mu}_{A_1} q'_1 & \mu \in (\Labels_1 \cup \{\tau\})\backslash \Labels_2} \]
\[  \infer{(q_1, q_2) \xrightarrow{\mu}_{A_{1 \parallel 2}} (q_1, q'_2)}{q_2 \xrightarrow{\mu}_{A_2} q'_2 & \mu \in (\Labels_2 \cup \{\tau\})\backslash \Labels_1} \]
\[  \infer{(q_1, q_2) \xrightarrow{\mu}_{A_{1 \parallel 2}} (q'_1, q'_2)}{q_1 \xrightarrow{\mu}_{A_1} q'_1 & q_2 \xrightarrow{\mu}_{A_2} q'_2  & \mu \in \Labels_1 \cap \Labels_2}\]
\end{definition}
The specification $S_{1} \parallel S_{2}$ shown in Fig.~\ref{fig:vending-spec-comp}~(a) represents the parallel composition of specifications 
$S_{1}$ and $S_{2}$.

\subsection{Hiding}
\label{sec:hiding}

The parallel composition of two components is often followed by {\em hiding} some of the actions on which 
they synchronize. We follow the process algebraic approach in which parallel composition and hiding 
operations are two separate operations, and formally define hiding as follows.

\begin{definition}[Hiding]
\label{def:hiding}
Let $A = (Q, \InputLabels, \OutputLabels, T, \qinit)$ be an $\iolts$ and $\Sigma \subseteq \OutputLabels$ be 
the subset of output actions. The {\em hiding} of $\Sigma$ in $A$, denoted by $\hide{A}{\Sigma}$, is the 
tuple $(Q, \InputLabels, \OutputLabels \setminus \Sigma, \hide{T}{\Sigma}, \qinit)$, where $\hide{T}{\Sigma}$ is 
obtained from $T$ by replacing every transition $(q,a,q') \in T$ labeled by an output action $a \in \Sigma$ by 
the transition $(q,\tau,q')$.
\end{definition}
The specification $\hide{S_{1} \parallel S_{2}}{\Sigma}$ shown in Fig.~\ref{fig:vending-spec-comp}~(b) represents the hiding of $\Sigma$ in  
$S_{1} \parallel S_{2}$, where $\Sigma = \{\mtee, \mcoffee, \mcoffeemilk, \done\}$.

\subsection{Input/Output Conformance Relation}
\label{sec:ioco}

Given an $\iolts$ $A$, the set $\out{q}{} \equiv  \{ a \in \OutputLabels~|~q \xrightarrow{a}~\} \cup \{\delta~|~\delta(q)\}$ is 
the set of all outputs (including $\delta$ if $q$ is quiescent) that are defined when the system is in state $q$. 
The set $q \; \after \; \sigma \equiv \{ q'~|~q \xRightarrow{\sigma}_{A_\delta} q'\}$ denotes the set of 
states that can be reached in $A$ from $q$ after reading $\sigma$ in its suspension automaton $A_\delta$. 
We now present the formal definition of the $\ioco$ relation.

\begin{definition}
\label{def:ioco}
Given a $\riolts$ $I$ and an 
$\iolts$ $S$, we say that 
$I \; \ioco \; S$ iff
$$
\forall \sigma \in \straces{S},\; \out{\qinit_{I} \; \after \; \sigma}{} \subseteq \out{\qinit_{S} \; \after \; \sigma}{}.
$$
\end{definition}

In the vending machine example, both $I_{1} \; \ioco \; S_{1}$ and $I_{2} \; \ioco \; S_{2}$, where $S_{1}$ and $S_{2}$ 
are depicted in Fig.~\ref{fig:vending-spec}~(a) and (b), and $I_{1}$ and $I_{2}$ are depicted in Fig.~\ref{fig:vending-impl}~(a) and 
(b), respectively. We now recall the results from~\cite{tretmans-comp} which state that $\ioco$ is not preserved  
in general under parallel composition and hiding, but is preserved if all the specifications are receptive. 
Receptiveness of specifications can be achieved by demonic completion (see~\cite{tretmans-comp} for its formal definition).

\begin{theorem}[\hspace{-1pt}\cite{tretmans-comp}]
Given two composable $\riolts$ $I_{1}$ and $I_{2}$, two composable $\iolts$ $S_{1}$ and $S_{2}$ and 
two composable $\riolts$ $S^{*}_{1}$ and $S^{*}_{2}$, we have
$$
\begin{array}{lcl}
I_{1} \; \ioco \; S_{1} \wedge I_{2} \; \ioco \; S_{2} &\not\rightarrow &(I_{1} \parallel I_{2}) \; \ioco \; (S_{1} \parallel S_{2}) \\
I_{1} \; \ioco \; S^{*}_{1} \wedge I_{2} \; \ioco \; S^{*}_{2} &\rightarrow &(I_{1} \parallel I_{2}) \; \ioco \; (S^{*}_{1} \parallel S^{*}_{2}).
\end{array}
$$
\end{theorem}

\begin{theorem}[\hspace{-1pt}\cite{tretmans-comp}]
Given a $\riolts$ $I$, an $\iolts$ $S$ and a $\riolts$ $S^{*}$, defined over the alphabet $\Labels$, and a subset $\Sigma \subseteq \OutputLabels$, we have
$$
\begin{array}{lcl}
I\ \ioco\ S &\not\rightarrow &\hide{I}{\Sigma}\ \ioco\ \hide{S}{\Sigma} \\
I\ \ioco\ S^{*} &\rightarrow &\hide{I}{\Sigma}\ \ioco\ \hide{S^{*}}{\Sigma}.
\end{array}
$$
\end{theorem}

Consider the vending machine example, in which $I_{1} \; \ioco \; S_{1}$ and $I_{2} \; \ioco \; S_{2}$, the sequence $\sigma = \gmoney \cdot \gutee$ and the compositions of specifications and implementations before and after hiding (see Fig.~\ref{fig:vending-spec-comp} 
and~\ref{fig:vending-impl-comp}). 
The available output after executing $\sigma$ on $I_{1} \parallel I_{2}$ is $\mcoffee$, while the composed specification 
$S_{1} \parallel S_{2}$ allows only $\delta$ after executing the same sequence, hence  $I_1 \parallel I_2$ does not $\ioco$-conform to $S_1 \parallel S_2$.
After hiding the actions $\Sigma = \{\mtee, \mcoffee, \mcoffeemilk, \done\}$ in the composition, 
we obtain additional traces which are not $\ioco$-conformant. 
For instance, after executing the sequence $\gmoney \cdot \gucoffee \cdot \gumilk$ in $\hide{I_{1}\parallel I_{2}}{\Sigma}$, the possible outputs are  
$\gcoffee$ and $\gcoffeemilk$, while the specification $\hide{S_{1} \parallel S_{2}}{\Sigma}$ allows only the action $\gcoffeemilk$ after 
executing the same sequence.

We note that demonic completion introduces new ``chaotic'' states to the original specification, from which 
all behaviors are allowed. Exploring chaotic states in test case generation is not useful. In 
order to avoid their exploration  
\cite{tretmans-comp} defines the set of $\textbf{Utraces}$.
Intuitively, $\textbf{Utraces}$ restricts $\textbf{Straces}$ by eliminating underspecified traces. 
We obtain the $\uioco$ conformance relation by replacing $\textbf{Straces}$ 
with $\textbf{Utraces}$ in the definition of $\ioco$. However, it turns out that $\uioco$ does not 
preserve compositional properties (see~\cite{tr} for details).

\section{Optimistic Approach to Composition and Hiding}
\label{sec:interface}

In this section, we formalize the {\em friendly} composition and hiding operations, presented informally in 
Section~\ref{sec:motivating}.

\subsection{Friendly Environments}
\label{sec:env}




We have seen in Section~\ref{sec:prelim} that 
parallel composition and hiding can introduce {\em ambiguous} states, and 
that $\ioco$ is not preserved under these two operations. 
An {\em ambiguous} state results from the parallel composition of two $\iolts$ in which one 
emits an action that the other one is not ready to accept.

\begin{definition}[Ambiguous state]
\label{def:error_state} 
Given two composable $\iolts$ $A$ and $B$, a pair $(q_A,q_B) \in Q_A \times Q_B$ is an {\em ambiguous state} 
if there exists a shared action $\alpha \in \Labels_A \cap \Labels_B$ such that either: 
(1) $\alpha\in \OutputLabels_A$, $q_A \xrightarrow{\alpha}{}$ and $q_B \not \xrightarrow{\alpha}{}$; or 
(2) $\alpha\in \OutputLabels_B$, $q_B \xrightarrow{\alpha}{}$ and $q_A \not \xrightarrow{\alpha}{}$. 
\end{definition}

In the parallel composition $S_{1} \parallel S_{2}$ of the vending machine example, depicted in 
Fig.~\ref{fig:vending-spec-comp}~(a), the state $3A$ is ambiguous because $S_{1}$ emits the output action  $\outmtee$, while $S_{2}$ does not accept it.



Inspired by contract-based design and interface theories, we 
propose an optimistic approach to composition and hiding. In this optimistic setting, we look for a {\em friendly environment}
which steers the specification away from ambiguous states. A friendly environment is helpful towards the systems, 
by always accepting the system's outputs and never providing actions that the system cannot accept as inputs.

\begin{definition}[Friendly environment]
\label{def:legal_env}
Given an $\iolts$ $A = (Q, \InputLabels, \OutputLabels, T, \qinit)$, a 
{\em friendly environment} for $A$ is a strongly-convergent $\iolts$ 
$E = (Q_{E}, \OutputLabels, \InputLabels, T_{E}, \qinit_{E})$ such that 
$A\parallel E$ does not have ambiguous states.
\end{definition}

A {\em composition-friendly} environment does not allow a composed system to reach an ambiguous state in the composition.

\begin{definition}[Composition-friendly environment]
\label{def:cfenv}
Given a composed $\iolts$ $A\parallel B$, a 
{\em composition-friendly environment} for $A \parallel B$ is its friendly environment such that 
for all ambiguous states $(q_A,q_B) \in Q_{A\parallel B}$, for all $q_E \in E$, $((q_A,q_B),q_E)$ is 
not reachable in $(A \parallel B) \parallel E$.
\end{definition}


A friendly environment is \emph{maximal} if it admits more behaviors than any other friendly environment.
The maximal friendly environment is used to compute the 
largest portion of the specification that is guaranteed to preserve conformance under composition and hiding. The resulting 
specification characterizes all sequences for which integration testing is not necessary. 

\begin{definition}[Maximal friendly environment]
\label{def:maxenv}
A friendly environment $E$ for an $\iolts$ $A$ is said to be {\em maximal} if for all 
friendly environments $E'$ for $A$ it holds $\traces{E'} \subseteq \traces{E}$.
\end{definition}


The $\iolts$ fragment that interacts correctly with its friendly environment $E$ is called its $E$-reachable fragment.
It is obtained by composing the $\iolts$ with $E$, while keeping the original meaning of inputs and outputs. 

\begin{definition}[Environment reachable fragment]
Let $A$ be an $\iolts$ and $E$ be its friendly environment. The \emph{$E$-reachable fragment of $A$} is an 
$\iolts$ $(Q, \InputLabels, \OutputLabels, T, \qinit)$, where $Q = Q_{A\parallel E}$, $\qinit = \qinit_{A\parallel E}$, $\InputLabels = \InputLabels_{A}$, $\OutputLabels = \OutputLabels_{A}$, and $T = T_{A\parallel E}$.
\end{definition}

\subsection{Friendly Composition, Friendly Hiding and $\ioco$}
\label{def:friendly}

We are now ready to formally define {\em friendly composition} and 
{\em hiding}, and show that $\ioco$ is a pre-congruence for these two operations.
 We use the maximal friendly
environment to restrict the classical parallel composition and hiding operations to a fragment that guarantees avoiding ambiguous 
states.

\begin{definition}[Friendly composition]
\label{def:int_comp}
Given two composable $\iolts$ $A$ and $B$, we say that they are {\em compatible} if there exists a   
composition-friendly environment $E$ for $A \parallel B$.
Given two compatible $\iolts$ $A$ and $B$, their {\em friendly composition}, denoted by $\intcomp{A}{B}$, is an 
$E$-reachable fragment of $A\parallel B$, where $E$ is the maximal composition-friendly environment for $A\parallel B$. 
\end{definition}

\begin{lemma}
For any two compatible $\iolts$ $A$ and $B$, there exists a maximal composition-friendly environment for $A \parallel B$.
\end{lemma}

\begin{definition}[Friendly hiding]
Given an $\iolts$ $A$ and $\Sigma \subseteq \OutputLabels$, 
the {\em friendly $\Sigma$-hiding} of $A$, 
denoted by $\fhide{A}{\Sigma}$, is an $E$-reachable fragment of $\hide{A}{\Sigma}$, where $E$ is the maximal 
friendly environment for $\hide{A}{\Sigma}$. 
\end{definition}
The specifications $S_{1} \otimes S_{2}$ and  $\fhide{S_{1} \otimes S_{2}}{\Sigma}$ depicted in Fig.~\ref{fig:vm-spec-intf}~(a) and 
(b) represent friendly composition of $S_{1}$ and $S_{2}$, followed by the friendly hiding of the synchronization actions 
$\Sigma = \{\mtee, \mcoffee, \mcoffeemilk, \done\}$.

We note that for receptive models, the friendly composition and friendly hiding coincide with the parallel composition and hiding. 
We state the main technical contributions of the paper --- that $\ioco$ relation is preserved under friendly composition and friendly hiding.

\begin{theorem}
\label{th:ioco_comp}
Given two compatible $\riolts$ $I_{1}$ and $I_{2}$ and two compatible $\iolts$ $S_{1}$ and $S_{2}$, we have
$$
I_{1} \; \ioco \; S_{1} \wedge I_{2} \; \ioco \; S_{2} \rightarrow (I_{1} \otimes I_{2}) \; \ioco \; (S_{1} \otimes S_{2}).
$$
\end{theorem}

\begin{theorem}
\label{th:ioco_hide}
Given a $\riolts$ $I$ and an $\iolts$ $S$ defined over the set $\OutputLabels$ of output labels and  $\Sigma \subseteq \OutputLabels$, we have
$$
I \; \ioco \; S \rightarrow \fhide{I}{\Sigma} \; \ioco \; \fhide{S}{\Sigma}.
$$
\end{theorem}

\begin{corollary}
\label{col:ioco_comp_hide}
Given two compatible $\riolts$ $I_{1}$ and $I_{2}$, two compatible $\iolts$ $S_{1}$ and $S_{2}$, and $\Sigma \subseteq \OutputLabels_{1\parallel 2}$, 
we have 
$$
I_{1} \; \ioco \; S_{1} \wedge I_{2} \; \ioco \; S_{2} \rightarrow \fhide{I_{1} \otimes I_{2}}{\Sigma} \; \ioco \; \fhide{S_{1} \otimes S_{2}}{\Sigma}.
$$
\end{corollary}

\subsection{Computing Friendly Composition and Hiding}
\label{sec:algo}

In this section, we present the algorithms for effectively computing the friendly composition and hiding.
We first create the {\em deterministic} maximal friendly environment $E$ for an $\iolts$ $A$. 
It is constructed by swapping input and output labels of $A$ and applying a variant of 
the subset construction that determinizes the $\iolts$ afterwards.

Our variant of the subset construction works as follows. Consider an $\iolts$ $A$, the standard subset construction $C$ of $A$, and a state $S$ of $C$. When $C$ is in $S$, the maximal environment must accept all output transitions from $S$. In contrast, the environment provides an input to $C$ only if all states in $S$ can accept it. As we can see, $E$ quantifies 
existentially over outputs in $A$ and universally over inputs in $A$. 



\begin{definition}[Maximal deterministic friendly environment]  
The {\em maximal deterministic friendly environment} of an $\iolts$ $A=(Q, \InputLabels, \OutputLabels, T, \qinit)$ is an 
$\iolts$  $\envdet{A} = (2^Q, \OutputLabels, \InputLabels, T_E, \hat{Q} )$, where 
$\hat{Q} = \qinit \; \after \; \epsilon$, and $(S, \alpha, S') \in T_E$ if: 1) $S' = S \; \after \; \alpha$; 2) $\alpha \in \InputLabels$ implies 
that $q \xrightarrow{\alpha}$ for all $q \in S$; and 3) $\alpha \in \OutputLabels$ implies that there exists $q \in S$ such that 
$q \xrightarrow{\alpha}$.
\end{definition}

\begin{algorithm}[h]
  \caption{Friendly composition}
  \label{alg:intcomp}
\begin{algorithmic}
\Require composable $\iolts$ $A_1$ and $A_2$
\Ensure $\intcomp{A_1}{A_2}$ or not compatible
\State $E \gets \envdet{A_1 \parallel A_2}$
\State $\emph{Amb} \gets$ states in $E$ that contain an ambiguous s.\ of $A_1 \parallel A_2$ 
\State $\emph{Prune} \gets \emptyset$
\ForAll{$S \in \emph{Amb}$}
	\State $\emph{Prune} \gets \emph{Prune} \;\cup\; \{\text{state $S'$ in $E$} \;|\;\exists \sigma\in {\OutputLabels_{A_1 \parallel A_2}}^*: S' \xRightarrow{ \sigma }_E S  \}$
\EndFor
\State remove states in $\emph{Prune}$ from $E$
\If {$E$ has no initial state}\ \Return not compatible
\Else\  \Return $E$-reachable fragment of $(A_1 \parallel A_2)$
\EndIf
\end{algorithmic}
\end{algorithm}

Algorithm \ref{alg:intcomp} constructs the friendly composition for $\iolts$ $A_1$ and $A_2$ or returns the information that they are not compatible. 
First, it constructs the maximal deterministic friendly environment $\envdet{A_1\parallel A_2}$. The algorithm then 
computes the set $\emph{Amb}$ of states 
in $\envdet{A_1\parallel A_2}$ that contain an ambiguous state from $A_1 \parallel A_2$ and prunes away all states in 
$\envdet{A_1\parallel A_2}$ that reach $\emph{Amb}$ by a trace of output labels in $A_1 \parallel A_2$. 
If the initial state is removed in the process, 
then no friendly environment for $A_1 \parallel A_2$ exists. Otherwise, $E$ is the resulting maximal composition-friendly environment for 
$A_1 \parallel A_2$ and the $E$-reachable fragment of $A_1 \parallel A_2$ is their friendly composition.
The friendly composition $\intcomp{A_1}{A_2}$ constructed by the algorithm is of size at most $\vert A_1 \vert \cdot \vert A_2 \vert \cdot 2^{\vert A_1 \vert \cdot \vert A_2 \vert}$ and if $A_1$ and $A_2$ are both deterministic, then $\intcomp{A_1}{A_2}$ is of size $\vert A_1 \vert \cdot \vert A_2 \vert.$

Algorithm \ref{alg:obs} computes the friendly $\Sigma$-hiding of an $\iolts$ $A$, where $\Sigma \subseteq \OutputLabels_A$. To obtain the maximal deterministic friendly environment, we simply hide actions in $A$, determinize it and then construct its maximal friendly environment $E$. 
The friendly hiding $\fhide{A}{\Sigma}$ computed by the algorithm is of size $2^{\vert A \vert}$ and if $A$ is deterministic, then $\fhide{A}{\Sigma}$ is of size $\vert A \vert$.
We note that there 
always exists a friendly environment $E$ for arbitrary $A$. It suffices that $E$ accepts all outputs from $A$ and does not provide any inputs. 


\begin{algorithm}[h]
  \caption{Friendly hiding}
  \label{alg:obs}
\begin{algorithmic}
\Require $\iolts$ $A$, $\Sigma \subseteq \OutputLabels_A$,
\Ensure $\fhide{A}{\Sigma}$
\State $E \gets \envdet{\hide{A}{\Sigma}}$
\State \Return $E$-reachable fragment of $\hide{A}{\Sigma}$
\end{algorithmic}
\end{algorithm}

\section{Evaluation}
\label{sec:eval}

To evaluate our approach to compositional testing in the $\ioco$-theory, we implemented a proof-of-concept tool for computing friendly composition and hiding. We applied the tool to the {\em alternating bit protocol} 
taken from the CADP 
examples repository \cite{CADP}. We were  interested in the quality of the component specifications with respect to compositional 
testing. We also compared the size of the friendly composition of the component specifications to the size of the specification obtained by the demonic completion approach.


The alternating bit protocol is a data transfer protocol supporting re-transmission of lost or corrupted messages. 
For the sake of simplicity, we assume a perfect link (no corruption of messages). The protocol consists of a sender $A$ and a receiver $B$.
Initially, the sender $A$ waits for data to be  transmitted (\emph{Put?}). Upon receiving the data, it sends it to $B$ together with a sequence bit (\emph{Data0!}) and waits for an acknowledgment (\emph{Ack0?}). If the acknowledgment does not arrive before a time-out,
$A$ re-transmits the data. This behavior is modeled by a $\tau$-transition, which abstracts away the actual time-out value. Once $A$
receives the acknowledgment, it flips the sequence bit and repeats the procedure (\emph{Data1!} and \emph{Ack1?}).


Receiver $B$ behaves in a similar way. It first waits for data marked with the sequence bit (\emph{Data0?}) and possibly takes a time-out transition. Upon receiving the data, it 
sends it to the external environment (\emph{Put!}) and, in a next step, sends an acknowledgment marked with the same sequence bit 
to the sender (\emph{Data0!}). The receiver then repeats the procedure with the flipped sequence bit.

\begin{figure}
\centering
\subfloat[][Sender $A$]{\raisebox{0.6cm}{\scalebox{0.5}{\begin{tikzpicture}[node distance=2.5cm,->,>=stealth', initial text=, thick]
\tikzstyle{state}=[circle, draw]
\tikzstyle{initstate}=[state,initial]
\tikzstyle{transition}=[->,>=stealth']
\definecolor{currentcolor}{rgb}{0.000,0.000,0.000}
\node [initstate, draw=currentcolor] (state0) {0};
\node [state, draw=currentcolor, above right of=state0] (state1) {1};
\node [state, draw=currentcolor, right of=state1] (state2) {2};
\node [state, draw=currentcolor, below right of=state0] (state3) {3};
\node [state, draw=currentcolor, right of=state3] (state4) {4};
\node [state, draw=currentcolor, above right of=state4] (state5) {5};

\path
	(state2) edge node[auto] {$\tau$} (state1)
			 edge [bend left] node[auto] {Ack1?} (state1)
			 edge node[auto] {Ack0?} (state5)
	(state3) edge node[auto] {$\tau$} (state4)
			 edge [bend left] node[auto] {Ack0?} (state4)
			 edge node[auto] {Ack1?} (state0)
	(state0) edge node[auto] {Put?} (state1)
	(state1) edge [bend left] node[auto] {Data0!} (state2)
	(state5) edge node[auto] {Put?} (state4)
	(state4) edge [bend left] node[auto] {Data1!} (state3);
\end{tikzpicture} }}}%
\qquad
\subfloat[][Receiver $B$]{\scalebox{0.5}{\begin{tikzpicture}[node distance=2.5cm,->,>=stealth', initial text=, thick]
\tikzstyle{state}=[circle, draw]
\tikzstyle{initstate}=[state,initial]
\tikzstyle{transition}=[->,>=stealth']
\definecolor{currentcolor}{rgb}{0.000,0.000,0.000}
\node [state, draw=currentcolor] (state2) {2};
\node [initstate, draw=currentcolor, below left of=state2] (state0) {0};
\node [state, draw=currentcolor, below of=state0] (state1) {1};
\node [state, draw=currentcolor, below right of=state2] (state5) {5};
\node [state, draw=currentcolor, below of=state5] (state4) {4};
\node [state, draw=currentcolor, below right of=state1] (state3) {3};

\path
	(state0) edge node[auto] {$\tau$} (state1)
			 edge [bend left] node[auto] {Data1?} (state1)
			 edge node[auto] {Data0?} (state2)
	(state4) edge node[auto] {$\tau$} (state5)
			 edge [bend left] node[auto] {Data0?} (state5)
			 edge node[auto] {Data1?} (state3)
	(state1) edge [bend left] node[auto] {Ack1!}  (state0)
	(state5) edge [bend left] node[auto] {Ack0!}  (state4)
	(state2) edge node[auto] {Received!} (state5)
	(state3) edge node[auto] {Received!} (state1);
\end{tikzpicture} }}%
\caption{CADP specifications of the sender and receiver component for a variant of the Alternating Bit-Protocol. 
The $\tau$ transitions model time-outs.}%
\label{fig:alternatingBitOriginal}%
\end{figure}
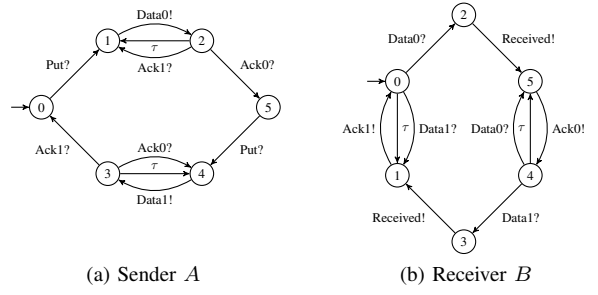

Sender $A$ and receiver $B$ (cf. Fig.~\ref{fig:alternatingBitOriginal}) are not compatible, i.e. no friendly environment guarantees the correctness of the 
protocol. There is a number of composition-ambiguous states in the parallel composition of $A$ and $B$, mainly due to the time-out 
($\tau$) transitions. For instance, $A$ can be either in state $1$ or $2$ after reading the trace $\emph{Put} \cdot \emph{Data0}$. In state 
$2$, $A$ expects the input $\emph{Ack0?}$ which is not the case in state $1$, where $A$ is ready to re-transmit $\emph{Data0!}$ and is brought to after a time-out. Similar problems occur with the receiver $B$ due to its own time-out transitions.

It follows that the specifications for $A$ and $B$ are too weak for compositional testing. In order to strengthen the 
specification of $A$, we need to improve the handling of the race between re-transmitting data to $B$ and receiving 
the acknowledgment from $B$. We tackle the problem by making the assumptions about the handling of an acknowledgment more explicit and introduce additional states in the specification. The similar time-out problem of receiver $B$ is handled in a slightly different way: time-out is no longer modeled as a $\tau$-transition but as a self-loop that allows $B$ continuous re-transmission of 
acknowledgments to $A$, while waiting for new data. The strengthened specifications $A'$ and $B'$ of the sender and receiver are depicted in 
Fig.~\ref{fig:alternatingBitFix}~(a) and (b). Their friendly composition (Fig.~\ref{fig:alternatingBitFix}~(c)) contains no composition-ambiguous state.

\begin{figure}
\centering
\subfloat[][Sender $A'$]{\raisebox{0.6cm}{\scalebox{0.5}{\begin{tikzpicture}[node distance=2.5cm,->,>=stealth', initial text=, thick]
\tikzstyle{state}=[circle, draw]
\tikzstyle{initstate}=[state,initial]
\tikzstyle{transition}=[->,>=stealth']
\definecolor{currentcolor}{rgb}{0.000,0.000,0.000}
\node [initstate, draw=currentcolor] (state0) {0};
\node [state, draw=currentcolor, above right of=state0] (state1) {1};
\node [state, draw=currentcolor, right of=state1] (state2) {2};
\node [state, draw=currentcolor, below right of=state0] (state3) {3};
\node [state, draw=currentcolor, right of=state3] (state4) {4};
\node [state, draw=currentcolor, above right of=state4] (state5) {5};
\node [state, draw=red, right of=state2] (state6) {6};
\node [state, draw=red, left of=state3] (state7) {7};
\path
	(state2) edge [draw=red] node[auto] {$\tau$} (state6)
			 edge [bend left] node[auto] {Ack1?} (state1)
			 edge node[auto] {Ack0?} (state5)
	(state3) edge [draw=red] node[auto] {$\tau$} (state7)
			 edge [bend left] node[auto] {Ack0?} (state4)
			 edge node[auto] {Ack1?} (state0)
	(state0) edge node[auto] {Put?} (state1)
			 edge[draw=red, loop right] node[auto, color=red] {Ack1?} (state0)
	(state1) edge [bend left] node[auto] {Data0!} (state2)
			 edge[draw=red, loop above] node[auto, color=red] {Ack1?} (state1)
	(state5) edge node[auto] {Put?} (state4)
			 edge[draw=red, loop left] node[auto, color=red] {Ack0?} (state5)
	(state4) edge [bend left] node[auto] {Data1!} (state3)
			 edge[draw=red, loop below] node[auto, color=red] {Ack0?} (state4)
	(state6) edge [draw=red] node[auto, color=red] {Ack0?} (state5)
			 edge [draw=red, bend right] node[color=red, above] {Data0!} (state2)
	(state7) edge [draw=red] node[auto, color=red] {Ack1?} (state0)
			 edge [draw=red, bend right] node[color=red, below] {Data1!} (state3);
\end{tikzpicture} }}}%
\hfill
\subfloat[][Receiver $B'$]{\scalebox{0.5}{\begin{tikzpicture}[node distance=2.5cm,->,>=stealth', initial text=, thick]
\tikzstyle{state}=[circle, draw]
\tikzstyle{initstate}=[state,initial right]
\tikzstyle{transition}=[->,>=stealth']
\definecolor{currentcolor}{rgb}{0.000,0.000,0.000}
\node [state, draw=currentcolor] (state2) {2};
\node [initstate, draw=currentcolor, below left of=state2] (state0) {0};
\node [state, draw=currentcolor, below of=state0] (state1) {1};
\node [state, draw=currentcolor, below right of=state2] (state5) {5};
\node [state, draw=currentcolor, below of=state5] (state4) {4};
\node [state, draw=currentcolor, below right of=state1] (state3) {3};

\path
	(state0) edge [draw=red, loop left] node[auto, color=red] {Ack1!} (state0)
			 edge [bend left] node[auto] {Data1?} (state1)
			 edge node[auto] {Data0?} (state2)
	(state4) edge [draw=red, loop right] node[auto, color=red] {Ack0!} (state5)
			 edge [bend left] node[auto] {Data0?} (state5)
			 edge node[auto] {Data1?} (state3)
	(state1) edge [bend left] node[auto] {Ack1!}  (state0)
			 edge [draw=red, loop left] node[auto, color=red] {Data1?} (state1)
	(state5) edge [bend left] node[auto] {Ack0!}  (state4)
			 edge [draw=red, loop right] node[auto, color=red] {Data0?} (state5)
	(state2) edge node[auto] {Received!} (state5)
			 edge [draw=red, loop below] node[auto, color=red] {Data0?} (state2)
	(state3) edge node[auto] {Received!} (state1)
			 edge [draw=red, loop above] node[auto, color=red] {Data1?} (state3);
\end{tikzpicture} }}%
\hfill
\subfloat[][$A' \otimes B'$]{\scalebox{0.5}{\begin{tikzpicture}[node distance=2.5cm,->,>=stealth', initial text=, thick]
\tikzstyle{state}=[circle, draw]
\tikzstyle{initstate}=[state,initial right]
\tikzstyle{transition}=[->,>=stealth']
\definecolor{currentcolor}{rgb}{0.000,0.000,0.000}
\node [initstate, draw=currentcolor] (state0) {0};
\node [state, draw=currentcolor, above right of=state0] (state1) {1};
\node [state, draw=currentcolor, right of=state1] (state2) {2};
\node [state, draw=currentcolor, right of=state2] (state3) {3};
\node [state, draw=currentcolor, above of=state2] (state4) {4};
\node [state, draw=currentcolor, right of=state4] (state5) {5};
\node [state, draw=currentcolor, below right of=state3] (state6) {6};
\node [state, draw=currentcolor, below left of=state6] (state7) {7};
\node [state, draw=currentcolor, left of=state7] (state8) {8};
\node [state, draw=currentcolor, left of=state8] (state9) {9};
\node [state, draw=currentcolor, below of=state8] (state10) {10};
\node [state, draw=currentcolor, left of=state10] (state11) {11};

\path
(state0)	edge[loop left] node[auto] {Ack1!}	(state0)
			edge node[auto] {Put?} 	(state1)
(state1)	edge[loop above] node[auto] {Ack1!} 	(state1)
			edge node[auto] {Data0!}	(state2)
(state2)	edge node[auto] {Received!} 	(state3)
			edge[bend right] node[auto] {$\tau$}	(state4)
(state4) 	edge node[auto] {Received!}	(state5)
			edge[bend right] node[left] {Data0!}	(state2)
(state3)	edge[bend right] node[auto] {$\tau$} (state5)
			edge node[left] {Ack0!}	(state6)
(state5)	edge[bend right] node[left] {Data0!}	(state3)
			edge[bend left] node[auto] {Ack0!}	(state6)
(state6) 	edge[loop right] node[auto] {Ack0!}	(state6)
			edge node[auto] {Put?} 	(state7)
(state7)	edge node[auto] {Data1!}	(state8)
			edge[loop below] node[auto] {Ack0!}	(state7)
(state8)	edge node[auto] {Received!}	(state9)
			edge[bend right] node[auto] {$\tau$}	(state10)
(state9)	edge node[right] {Ack1!}	(state0)
			edge[bend right] node[auto] {$\tau$}	(state11)
(state10)	edge node[auto] {Received!}	(state11)
			edge[bend right] node[right] {Data1!}	(state8)
(state11)	edge[bend left] node[auto] {Ack1!}	(state0)
			edge[bend right] node[right] {Data1!}	(state9);
\end{tikzpicture} }}%
\caption{Strengthened specifications (a) $A'$; (b) $B'$; and (c) their friendly composition.}%
\label{fig:alternatingBitFix}%
\end{figure}

Hiding all synchronization actions ($\emph{Data}$ and $\emph{Ack}$) in $A' \otimes B'$ introduces new ambiguous states. 
A friendly environment cannot observe the internal state of $A' \otimes B'$ and decide when the protocol is ready to receive new data 
items ($\emph{Put?}$ action). The easy way to overcome this problem is to add constraints to the original specification of 
the sender that say how to handle the input $\emph{Put?}$ when the sender is in a non-observable state. The solution we use in this example, however, is 
to strengthen the sender specification by adding an output action $\emph{Ready!}$ which tells the external environment that it is ready to 
accept new data. The resulting specification $A''$ is depicted in Fig.~\ref{fig:alternatingBitHiding}~(a). The new specification 
requires a hand-shake between the protocol and the environment and results in the friendly composition 
$\fhide{A'' \otimes B'}{\Sigma}$ followed by the friendly hiding of $\Sigma = \{ \emph{Ack0}, \emph{Ack1}, \emph{Data0}, \emph{Data1}\}$. 
The composite specification $\fhide{A'' \otimes B'}{\Sigma}$ does not encounter any ambiguous states. It follows that any
implementation of sender $A''$ and receiver $B'$ can be tested individually and that their composition is correct-by-construction, 
without need for additional integration tests.

\begin{figure}
\centering
\subfloat[][Sender $A''$]{\raisebox{0.5cm}{\scalebox{0.5}{\begin{tikzpicture}[node distance=2.5cm,->,>=stealth', initial text=, thick]
\tikzstyle{state}=[circle, draw]
\tikzstyle{initstate}=[state,initial]
\tikzstyle{transition}=[->,>=stealth']
\definecolor{currentcolor}{rgb}{0.000,0.000,0.000}
\node [initstate, draw=currentcolor] (state0) {0};
\node [state, draw=currentcolor, above right of=state0] (state1) {1};
\node [state, draw=currentcolor, right of=state1] (state2) {2};
\node [state, draw=currentcolor, below right of=state0] (state3) {3};
\node [state, draw=currentcolor, right of=state3] (state4) {4};
\node [state, draw=currentcolor, above right of=state4] (state5) {5};
\node [state, draw=red, right of=state2] (state6) {6};
\node [state, draw=red, left of=state3] (state7) {7};
\node [state, draw=red, left of=state1] (state8) {8};
\node [state, draw=red, right of=state4] (state9) {9};
\path
	(state2) edge [draw=red] node[auto] {$\tau$} (state6)
			 edge [bend left] node[auto] {Ack1?} (state1)
			 edge node[auto] {Ack0?} (state5)
	(state3) edge [draw=red] node[auto] {$\tau$} (state7)
			 edge [bend left] node[auto] {Ack0?} (state4)
			 edge node[auto] {Ack1?} (state0)
	(state0) edge [draw=red] node[auto, color=red] {Ready!} (state8)
			 edge[draw=red, loop right] node[auto, color=red] {Ack1?} (state0)
	(state1) edge [bend left] node[auto] {Data0!} (state2)
			 edge[draw=red, loop above] node[auto, color=red] {Ack1?} (state1)
	(state5) edge[draw=red] node[auto, color=red] {Ready!} (state9)
			 edge[draw=red, loop left] node[auto, color=red] {Ack0?} (state5)
	(state4) edge [bend left] node[auto] {Data1!} (state3)
			 edge[draw=red, loop below] node[auto, color=red] {Ack0?} (state4)
	(state6) edge [draw=red] node[auto, color=red] {Ack0?} (state5)
			 edge [draw=red, bend right] node[color=red, above] {Data0!} (state2)
	(state7) edge [draw=red] node[auto, color=red] {Ack1?} (state0)
			 edge [draw=red, bend right] node[color=red, below] {Data1!} (state3)
	(state8) edge [draw=red,loop above] node [auto, color=red] {Ack1?} (state8)
	         edge node[auto] {Put?} (state1)
	(state9) edge [draw=red,loop below] node [auto, color=red] {Ack0?} (state9)
	         edge node[auto] {Put?} (state4);
\end{tikzpicture} }}}%
\hfill
\subfloat[][Receiver $B'$]{\scalebox{0.5}{\begin{tikzpicture}[node distance=2.5cm,->,>=stealth', initial text=, thick]
\tikzstyle{state}=[circle, draw]
\tikzstyle{initstate}=[state,initial right]
\tikzstyle{transition}=[->,>=stealth']
\definecolor{currentcolor}{rgb}{0.000,0.000,0.000}
\node [state, draw=currentcolor] (state2) {2};
\node [initstate, draw=currentcolor, below left of=state2] (state0) {0};
\node [state, draw=currentcolor, below of=state0] (state1) {1};
\node [state, draw=currentcolor, below right of=state2] (state5) {5};
\node [state, draw=currentcolor, below of=state5] (state4) {4};
\node [state, draw=currentcolor, below right of=state1] (state3) {3};

\path
	(state0) edge [draw=red, loop left] node[auto, color=red] {Ack1!} (state0)
			 edge [bend left] node[auto] {Data1?} (state1)
			 edge node[auto] {Data0?} (state2)
	(state4) edge [draw=red, loop right] node[auto, color=red] {Ack0!} (state5)
			 edge [bend left] node[auto] {Data0?} (state5)
			 edge node[auto] {Data1?} (state3)
	(state1) edge [bend left] node[auto] {Ack1!}  (state0)
			 edge [draw=red, loop left] node[auto, color=red] {Data1?} (state1)
	(state5) edge [bend left] node[auto] {Ack0!}  (state4)
			 edge [draw=red, loop right] node[auto, color=red] {Data0?} (state5)
	(state2) edge node[auto] {Received!} (state5)
			 edge [draw=red, loop below] node[auto, color=red] {Data0?} (state2)
	(state3) edge node[auto] {Received!} (state1)
			 edge [draw=red, loop above] node[auto, color=red] {Data1?} (state3);
\end{tikzpicture} }}%
\hfill
\subfloat[][$\fhide{A'' \otimes B'}{\Sigma}$]{\scalebox{0.5}{\begin{tikzpicture}[node distance=2.5cm,->,>=stealth', initial text=, thick]
\tikzstyle{state}=[circle, draw]
\tikzstyle{initstate}=[state,initial right]
\tikzstyle{transition}=[->,>=stealth']
\definecolor{currentcolor}{rgb}{0.000,0.000,0.000}
\node [initstate, draw=currentcolor] (state0) {0};
\node [state, draw=currentcolor, above right of=state0] (state1) {1};
\node [state, draw=currentcolor, right of=state1] (state2) {2};
\node [state, draw=currentcolor, right of=state2] (state3) {3};
\node [state, draw=currentcolor, above of=state2] (state4) {4};
\node [state, draw=currentcolor, right of=state4] (state5) {5};
\node [state, draw=currentcolor, below right of=state3] (state6) {6};
\node [state, draw=currentcolor, below left of=state6] (state7) {7};
\node [state, draw=currentcolor, left of=state7] (state8) {8};
\node [state, draw=currentcolor, left of=state8] (state9) {9};
\node [state, draw=currentcolor, below of=state8] (state10) {10};
\node [state, draw=currentcolor, left of=state10] (state11) {11};
\node [state, draw=currentcolor, left of=state1] (state12) {12};
\node [state, draw=currentcolor, right of=state7] (state13) {13};

\path
(state0)	edge[loop left] node[auto] {$\tau$}	(state0)
			edge node[auto] {Ready!} 	(state12)
(state1)	edge[loop above] node[auto] {$\tau$} 	(state1)
			edge node[auto] {$\tau$}	(state2)
(state2)	edge node[auto] {Received!} 	(state3)
			edge[bend right] node[auto] {$\tau$}	(state4)
(state4) 	edge node[auto] {Received!}	(state5)
			edge[bend right] node[left] {$\tau$}	(state2)
(state3)	edge[bend right] node[auto] {$\tau$} (state5)
			edge node[left] {$\tau$}	(state6)
(state5)	edge[bend right] node[left] {$\tau$}	(state3)
			edge[bend left] node[auto] {$\tau$}	(state6)
(state6) 	edge[loop right] node[auto] {$\tau$}	(state6)
			edge node[auto] {Ready!} 	(state13)
(state7)	edge node[auto] {$\tau$}	(state8)
			edge[loop below] node[auto] {$\tau$}	(state7)
(state8)	edge node[auto] {Received!}	(state9)
			edge[bend right] node[auto] {$\tau$}	(state10)
(state9)	edge node[right] {$\tau$}	(state0)
			edge[bend right] node[auto] {$\tau$}	(state11)
(state10)	edge node[auto] {Received!}	(state11)
			edge[bend right] node[right] {$\tau$}	(state8)
(state11)	edge[bend left] node[auto] {$\tau$}	(state0)
			edge[bend right] node[right] {$\tau$}	(state9)
(state12)
			edge node[auto] {Put?} 	(state1)
			edge[loop above] node[auto] {$\tau$} (state12)
(state13)
			edge node[auto] {Put?} 	(state7)
			edge[loop below] node[auto] {$\tau$} (state13);
\end{tikzpicture} }}%
\caption{Strengthened specifications (a) $A''$; (b) $B'$; and (c) their friendly composition followed by 
friendly hiding of $\Sigma$.}%
\label{fig:alternatingBitHiding}%
\end{figure}

We finally compare the size of $\fhide{A'' \otimes B'}{\Sigma}$ to the one of $\hide{d(A)\parallel d(B)}{\Sigma}$, where 
$d(A)$ and $d(B)$ denote the demonically completed variants of $A$ and $B$. The 
results are shown in Table~\ref{tab:results}. We first observe that by applying our approach we obtain specifications of 
smaller size than by demonically completing the component specifications and then applying parallel composition and 
hiding. This is in particular visible when comparing the size of $\hide{d(A) \parallel d(B)}{\Sigma}$ 
($76$ transitions and $34$ states), to the one of $\fhide{A'' \otimes B'}{\Sigma}$ ($24$ transitions and $12$ states). 
We note that in our approach, only foreseen interactions between components are taken into account, which is not 
the case with the demonic completion approach. While our framework may require manual improvement of the 
component specifications, we argue that this is the right procedure to arrive at specifications of good quality for 
compositional testing. Although more automated, the demonic completion approach to compositional testing  
admits many useless implementations.

\begin{table}
\centering
\begin{tabular}{|c|c|c|}
\hline 
\rule[-1ex]{0pt}{2.5ex} Model & \# tran & \# states \\ 
\hline 
\hline
\rule[-1ex]{0pt}{2.5ex} $A$ & 10 & 6 \\ 
\hline 
\rule[-1ex]{0pt}{2.5ex} $B$ & 10 & 6 \\ 
\hline 
\rule[-1ex]{0pt}{2.5ex} $d(A)$ & 31 & 9 \\ 
\hline 
\rule[-1ex]{0pt}{2.5ex} $d(B)$ & 28 & 9 \\ 
\hline
\rule[-1ex]{0pt}{2.5ex} $A''$ & 22 & 10 \\
\hline
\rule[-1ex]{0pt}{2.5ex} $B'$ & 14 & 6 \\ 
\hline
\hline
\rule[-1ex]{0pt}{2.5ex} $\hide{d(A) \parallel d(B)}{\Sigma}$ & 76 & 34 \\ 
\hline 
\rule[-1ex]{0pt}{2.5ex} $\fhide{A'' \otimes B'}{\Sigma}$ & 24 & 12 \\ 
\hline 
\end{tabular} 
\caption{Sizes of specification models and their compositions.}
\label{tab:results}
\vspace{-24pt}
\end{table}

To summarize, the case study shows that the alternating bit protocol specification was not modeled with compositional testing in mind.
Parts of the sender and receiver specifications are not sufficiently specified for cooperative interactions and do not admit 
compositional testing with $\ioco$. We improved the specifications by strengthening the assumptions where needed. We note that 
despite the strengthening, the improved specifications are not input-enabled. 

We finally remark that although the individual component specifications 
are strongly convergent, their composition with hiding is not. This may be a problem for testing 
with quiescence in general (see~\cite{tretmans-mbt}) but does not affect our work as presented here: we only require friendly environments to be strongly-convergent.

\section{Related Work}
\label{sec:related}

This paper is inspired by~\cite{tretmans-comp} and extends it by defining new composition and hiding 
operations adapted for under-specified models and preserving compositional properties of $\ioco$. 
Compositional properties of the real-time conformance relation $\tioco$ were studied in~\cite{tripakis-testing}. 
In order to preserve compositional testing with $\tioco$, specifications are required to be receptive.
Compositional properties of the $\cspio$ conformance relation for model-based testing with CSP specifications were studied in~\cite{cspio}. 
CSP operations are shown to be monotonic with respect to $\cspio$ when the specifications are input-enabled, 
or the implementations are not receptive, and after each trace, the input actions accepted by the implementation 
are a subset of those offered by the specification.  The compositional testing problem 
for systems modeled as networks of abstract components, based on coalgebraic definitions, was considered in~\cite{aiguier}.
Once again, specifications must be receptive to preserve compositional properties in testing.
Assume-guarantee reasoning is combined with 
$\ioco$ in~\cite{ag-ioco} in order to allow compositional testing. This work is complementary to ours, as it starts from a 
global specification of the complete system, and uses assumptions about components to divide and conquer the testing process. 
A methodology to reduce the efforts of integration testing is presented in~\cite{mbi}. It combines model-based integration with model-based testing, 
but does not provide formal arguments that support the proposed approach.

This paper is also inspired by the interface theories~\cite{intf1,intf2}. In contrast 
to interface automata, used in the context of contract-based design, this work focuses on compositional properties in testing. 
Instead of iterative design through stepwise refinement, the $\ioco$-theory assumes the existence of an implementation. 
We consider $\ioco$ with its explicit treatment of quiescence as the refinement relation, rather than alternation simulation used in interface 
automata. The integration of specifications in the $\ioco$-theory separates parallel composition from hiding, thus allowing for multicast and broadcast communication. In interface automata, parallel composition and 
hiding are combined into a single operation, thus allowing point-to-point communication only. We also mention similar frameworks 
in contract-based design: synchronous interfaces with and without shared variables~\cite{shared}, synchronous relational interfaces~\cite{relational}, and real-time interfaces~\cite{kim-timed,tom-timed}.
\section{Conclusion and Future Perspectives}
\label{sec:conclusion}

We proposed a novel approach to compositional testing for $\ioco$ based 
on friendly composition and hiding. Our framework characterizes foreseen interactions between components and 
minimizes the effort needed for integration testing. In addition to the technical results, this paper 
gives new insights to compositional testing in general and the associated difficulties. In particular, we use 
our approach to provide guidelines for identifying weaknesses in component specifications and improving them 
with compositional testing in mind.  
Since high-level specifications are typically much smaller than the actual implementations, we  
argue that this additional effort in model analysis is rewarded with a reduction of effort in expensive 
integration testing.

In our framework, we assume that the composition of component specifications is the specification of the 
integrated system. In the future, we will study how to exploit our results when the overall system is specified with 
a separate model. In particular, we will investigate whether our results can be combined with~\cite{ag-ioco}. 
In addition, we will study whether we can weaken the notion of the ambiguous states, while preserving 
compositional properties of $\ioco$. Although we considered asynchronous models and $\ioco$-theory, 
we are confident that our results can be adapted to different modeling frameworks and conformance relations. 
In fact, many issues related to compositional testing come from the power of the $\iolts$ model, 
where components compete in executing the actions without much restrictions. We will adapt our work to 
the synchronous data-flow systems, which we believe have more robust properties with respect to 
composition and hiding.
\section*{Acknowledgments}
\label{acknow}
This research was funded in part by the European Research Council (ERC) under grant agreement 267989
(QUAREM), by the ARTEMIS JU under grant agreement numbers 269335 (MBAT) and 295373 (nSafeCer), 
and by the Austrian Science Fund (FWF) project S11402-N23 (RiSE).

\bibliographystyle{IEEEtran}
\bibliography{ref}
%




\end{document}